\documentclass[12pt,a4paper,final, twocolumn]{aastex63}
\usepackage{graphicx}
\usepackage{hyperref}
\usepackage{amsmath}
\usepackage{natbib}
\usepackage{aas_macros}
\begin{document}

\title[]{Data challenges as a tool for time-domain astronomy}

\author{Ren\'ee Hlo\v{z}ek}
\address{Department of Astronomy and Astrophysics, University of Toronto ON M5S 3H4, Canada: \url{hlozek@dunlap.utoronto.ca}}
\address{Dunlap Institute for Astronomy and Astrophysics, University of Toronto ON M5S 3H4, Canada}

\begin{abstract}
Data challenges are emerging as powerful tools with which to answer fundamental astronomical questions. Time-domain astronomy lends itself to data challenges, particularly in the era of classification and anomaly detection. With improved sensitivity of wide-field surveys in optical and radio wavelengths from surveys like the Large Synoptic Survey Telescope (LSST) and the Canadian Hydrogen Intensity Mapping Experiment (CHIME), we are entering the large-volume era of transient astronomy. We highlight some recent time-domain challenges, with particular focus on the Photometric LSST Astronomical Time series Classification Challenge (PLAsTiCC), and describe metrics used to evaluate the performance of those entering data challenges.
\end{abstract}

\pacs{00.00, 20.00, 42.10}
\vspace{2pc}

\section{Introduction}
Time-domain astronomy has a rich history. From the first detection of the guests or `novae' in the sky, repeated observations of how the sky changes yields ever-increasing data sets of objects that brighten and fade once, so-called `transient' events or by changing brightness repeatedly as variable objects.

Combining data across different wavelength bands enables astronomers to learn about the underlying physical mechanisms for the emission of light, and the properties of these bright and changing visitors in the sky. However, gathering data using this multi-wavelength approach means that the final data set that results is heterogeneous in nature. Surveys are designed with an observational strategy, often referred to by the `cadence' of observations of different sky patches that serves many different science cases. As a result, a given patch of sky can be observed with repeat observations separated from hours to days. When adding in additional complications like weather at the observing site, and whether a particular region of sky will be visible at certain times, the final data that result are never evenly sampled in time. Both the multi-wavelength and non-uniform sampling of the data make astronomical time-series data rich and complex.

Data rates in astronomy are increasing as more advanced and `faster' surveys and instruments with wide-field capabilities come online. Surveys like the Catalina Real-time Transient Survey \citep[CRTS,][]{2012IAUS..285..306D}, the Zwicky Transient Facility \citep[ZTF,][]{2014htu..conf...27B}, Panoramic Survey Telescope and Rapid Response System \citep[PAN-STARRS1,][]{2016arXiv161205560C} are delivering large numbers of transient `alerts' to their collaborators, with alerts and triggers shared publicly for particular events of interest, via astronomers telegrams.  The collaborations above have a focus on astronomical transients, such as supernovae and stellar explosions, and variable stars. Other missions like the Lincoln Near-Earth Asteroid Research \citep[LINEAR,][]{2015arXiv150502082P} and the NASA Jet Propulsion Lab mission for  Near-Earth Asteroid Tracking \citep[NEAT,][]{1998BAAS...30.1036H} focus on the monitoring of near-earth objects (NEOs) and provided an impressive catalog of confirmed detections of NEOs, while institutions like the Japanese Spaceguard Association\footnote{\url{http://www.spaceguard.or.jp/ja/index.html}} continue to monitor the sky in optical and infrared light with the same goal.

The next generation Large Synoptic Survey Telescope \citep[LSST,][]{2002SPIE.4836...10T}, will generate orders of magnitude more transient alerts than previous surveys. With projections of around $10^7$ alerts per night\footnote{The expected specifications for survey parameters, specifications and requirements LSST are summarized online: \url{https://www.lsst.org/scientists/keynumbers}}, the need to develop alert stream `brokers' that will act as intermediaries, triggering follow-up telescopes is proving a critical and exciting challenge for the astronomy community.
The increase in data volume for these new astronomical surveys impressive: petabytes of data are expected for LSST alone. 

The challenge of large heterogeneous data is not unique to optical surveys. The Canadian Hydrogen Intensity Mapping \citep[CHIME,][]{2018ApJ...863...48C} will generate 2 PetaBytes of data per year, much of this data will be filtered to search for transients and radio bursts in real time. Similarly, the Square Kilometer Array \citep[SKA,][]{2018arXiv181102743S} and its precursors MeerKAT \cite{2017arXiv171104132F} and the the Australian SKA Pathfinder \citep[ASKAP, ][]{2009IEEEP..97.1507D} will generate large numbers of transients in the radio image plane.
These large data sets make astronomy an ideal test-bed for new methodologies and statistical approaches to signal and image processing of time-series data.

Time-domain science was recognized as a critical path of scientific inquiry in the 2010 National Research Councils Decadal Survey of Astronomy and Astrophysics, which resulted in significant investment in LSST for optical/IR science. 
The coming decade might well be known as the new golden age of transient astronomy and its related multi-wavelength synergies. 

In this review, I will set the stage for data challenges in astronomy generally, with a focus on time-series challenges as a tool to tackle the data problems in astronomy. I will discuss metrics for evaluating such time-domain challenges and finally discuss some recent efforts in this field. This new area of research has generated partnerships across scientific and computational fields such as computer science, statistics and machine learning. Combining these fields with traditional astronomy areas of interest has led to the formation of a new and active sub-field of astro-informatics.

\section{Data Challenges in Astronomy}

A data challenge is where a real or simulated data set is released to the community with the goal of testing different methods in the community to either derive a product, such as the lensing shear of a group of galaxies, or to classify an object into different classes such as whether or not an object is a strongly lensed system.

The idea of a `challenge' on a common set of data stems from the growth in the size of data sets, or the complexity of the problem at hand, and the need for reproducibility of results between different groups of researchers.

Astronomy has long utilized data challenges to focus the community around complex computational or methodological issues, and as a tool for gathering new solutions to known problems. As an example, in the field of gravitational lensing, challenges range from those requiring participants to develop the best lensing shear estimators in the series of GRavitational lEnsing Accuracy Testing challenges  \citep[GREAT,][]{great3}, to ones focused on identifying strong gravitational lens systems \cite{2018arXiv180203609M}. These challenges have typically focused on extracting information from astronomical images, and exploited interest from astronomers with their domain knowledge, and those interested in the problem purely from a computational stand point. Other challenges like the Radial Velocity Challenge \cite{rvresults} simulated radial velocity signals \cite{rvsims} from the then future planet-hunting missions like the Transiting Exoplanet Survey Satellite \citep[TESS][]{tess} to test and compare different community models of exoplanet discovery.

\begin{figure*}[htbp!]
    \centering
    \includegraphics[width=0.95\textwidth]{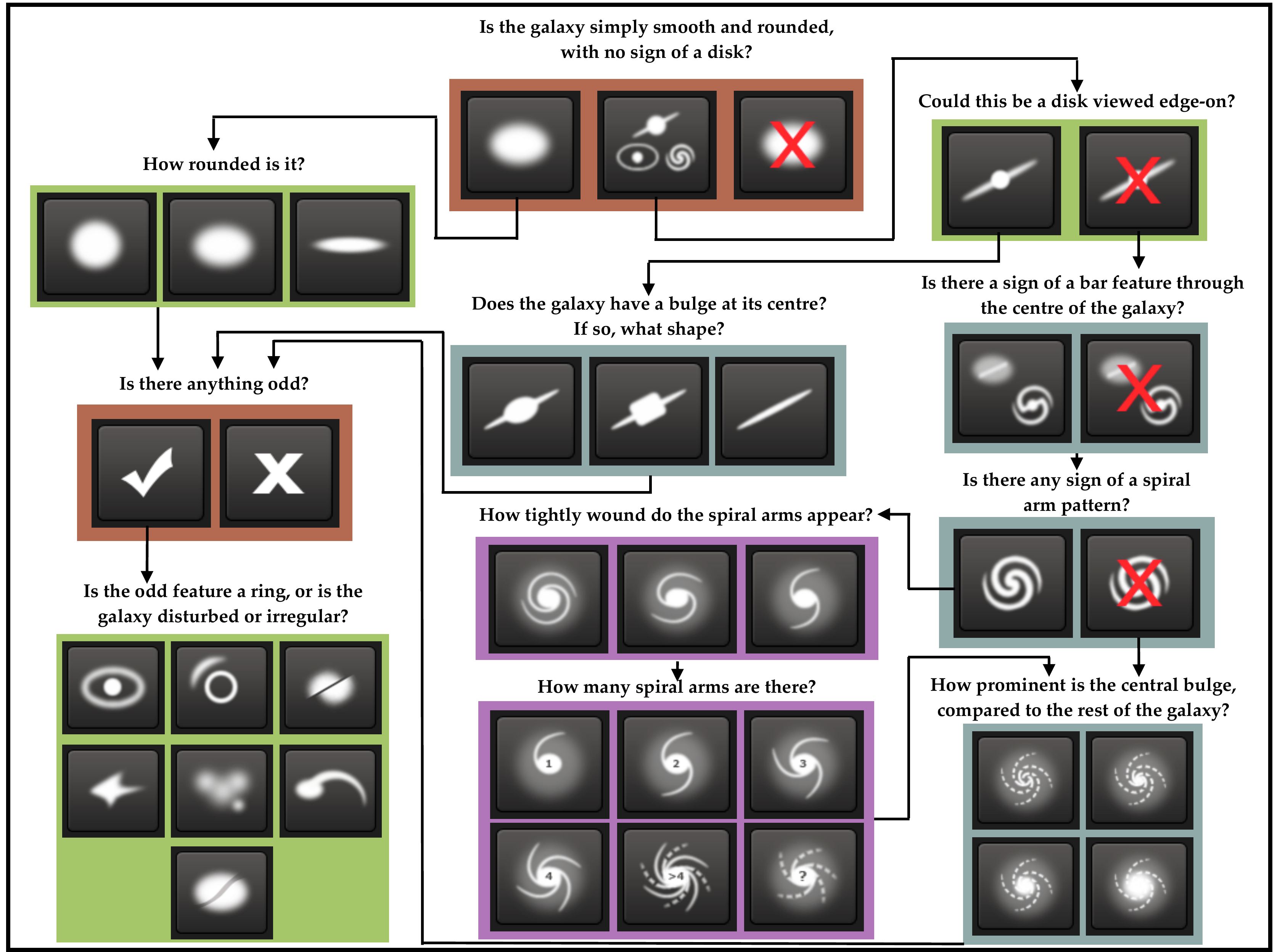}
    \caption{Citizen science classification flowchart from Galaxy Zoo. The success of the Galaxy Zoo and the series of classification challenges it generated under the umbrella of the `Zooniverse' has enabled scientists to answer known questions and has uncovered surprising new objects. Figure reproduced with permission from \cite{2013MNRAS.435.2835W}. 
    \label{fig:galzoo}}
\end{figure*}

In addition to challenges aimed at the expert astronomy or computer science communities directly, some challenges have utilized the power of citizen science to answer previously identified questions from established surveys. These challenges often result in new avenues of discovery that arise from data exploration by the citizen scientists. The discovery of `green pea' galaxies, which are aptly named to reflect the size and greenish appearance in the Sloan Digital Sky Survey images that were processed by citizen scientists as part of the Galaxy Zoo project\footnote{\url{https://www.zooniverse.org/}} has led to 22 additional publications on the metal-poor galaxies and their role in the reionization of the universe \cite{2009MNRAS.399.1191C}. Figure \ref{fig:galzoo} shows the decision tree that citizen scientists were required to use as part of the Galaxy Zoo classification efforts. The first layer separated classifications into elliptical and spiral-like categories, with sub-questions relating to the specifics of the now separate branches. This hierarchical classification scheme was central to the success of the Galaxy, as it was designed so that morphological questions appeared in increasing complexity as the structure was being identified \cite{2017MNRAS.464.4176W}.

Other established surveys have exploited similar partnerships with citizen science platforms like the Zooniverse to great effect, such as the Pan-STARRS survey which formed the challenge ``Supernova hunters."\footnote{\url{https://www.zooniverse.org/projects/dwright04/supernova-hunters}} Participants in the supernova hunters challenge have to date performed over 80 000 classifications of difference images formed from the subtraction of an image taken on one night from another image taken previously with possible transients in them.
Similar challenges by other groups, like NASA's Near Earth Orbit Hunter\footnote{\url{https://urlzs.com/PJyxV}} provided online ``credits'' for successful detections from a multi-facility data set, and used the system to provide incentives for public participation. 

While some challenges are designed to explore known data sets, others are designed to prepare users for upcoming data. The Laser Interferometer Space Antenna (LISA) data challenge\footnote{\url{https://lisa-ldc.lal.in2p3.fr/ldc}} \cite{2017arXiv170200786A} built upon the interest surrounding the recent discovery of gravitational wave sources and their electromagnetic counterparts \cite{2015arXiv150502082P} to focus the community on extracting signals from noisy data. Similarly, the ASKAP group designed a data challenge to test the fidelity of radio source finding \cite{2015PASA...32...37H} to ensure that various detection codes are well tested before the full onslaught of the data.

Time-series catalog-based (rather than image-based) challenges have also had great success within the transient community. The first large-scale classification challenge specific to supernova data was the Supernova Photometric Classification Challenge \citep[SNPhotCC,][]{2010arXiv1001.5210K}, which ran for four months in early 2010 and presented a blended mix of simulated supernova types, from a combination of SNIa models and a few core-collapse supernova models.  The simulation was generated using survey parameters similar to those of the Dark Energy Survey\footnote{\url{https://www.darkenergysurvey.org/}} \citep[see][for the recent supernova cosmology results from the survey]{2018arXiv181109565D}. 

The SNPhotCC released both a training and test data set to the broader astronomical community. After the challenge was complete, the truth table was also released publicly. While the main challenge lasted a few months \cite{2010PASP..122.1415K}, the data proved useful to the community long after the official end date of the challenge, and inspired many approaches to supernova classification. The main issues that surfaced during the challenge were the issues of training data non-representativity and potential classification bias. SNPhotCC was mainly focused on ensuring high purity for the resultant SNIa sample, as its main goal was to obtain a sample of SNeIa that were useful for photometric supernova cosmology without spectroscopic confirmation. Given the increased depth of LSST, the natural next step in simulating classification challenges from photometry is to simulate the LSST sky. With support from the LSST Corporation, a group was formed to develop the Photometric LSST Astronomical Time-series Classification Challenge \citep[PLAsTiCC,][]{2018arXiv181000001T}.  

The goals PLAsTiCC were to produce a more complete challenge by simulating a wider range of transients and variable objects, and to create a realization of an LSST-like survey. 
Wide-field surveys generate large volumes of data, and surveys that push to greater magnitude limits. Due to the selection bias of surveys, this means that much of the new data taken from such a deep survey has different properties as a function of redshift, or depth than any data from shallower surveys, and as such data used to train models will not be representative samples of the final data from the deeper survey. LSST will achieve both wide sky coverage and increased depth. As such, PLAsTiCC was designed to include both with large volumes of data and to be non-representative. The training data contained 8000 objects while the test data contains closer of three million objects. Moreover, the challenge that was generated with the non-astronomer in mind, the ethos of PLAsTiCC was that it does not rely on astronomical domain knowledge in order to participate. The challenge is hosted in partnership with the  Kaggle platform\footnote{\url{https://kaggle.com/c/PLAsTiCC-2018}}. Kaggle is a subsidiary company of Google with a mandate to host and facilitate data-intensive challenges. The preparation of PLAsTiCC started with an open call to the astronomy community to submit relevant models of transients and variable objects. These models were released to the community at the close of the challenge \cite{2019arXiv190311756K}. By the end of the challenge 1085 teams had been formed to participate in PLAsTiCC, evidence of the potential of such data challenges to capture the imagination of people interested in astronomical data science.
In all the data challenge examples highlighted above, the needs of current and future surveys and data sets informed the design of the challenge. The challenges themselves, however, have typically yielded new and surprising results and challenge techniques beyond the initial expectations of those initiating the data challenges, making them an attractive tool for time-domain astronomy.


\section{Time-domain challenge outcomes}
Data challenges in time-domain astronomy can typically be separated into two groups depending on the motive behind the challenge: some challenges require participants to classify objects into different ``known,'' or ``learned'' classes given a training set, while other challenges are focused on identifying anomalies in a given data set, which can be performed in unsupervised learning frameworks, without the need for training data.

We now discuss examples of both types of time-series challenges.
\subsection{Classification} 
Classification challenges are often designed with a focus on correctly identifying one type of class from the data set. For example, one might be focused on building a `clean' sample of Type Ia Supernovae in order to facilitate cosmological analyses. As such, the purity of the resultant sample, or the number of correct type classifications of a given object relative to the positive classifications of the entire sample, becomes an important metric in evaluating entries to the challenge. This is particularly relevant if it is costly to confirm the type of an object (e.g. by taking a spectrum with a ground-based instrument). Conversely, one might want to build a sample that is as complete in one type/class as possible, forgiving some bias from interlopers. In that case, the efficiency, or the number of true positive (TP) classifications relative to the total sum of classifications of that type, either TP or false negative (FN) classifications, becomes the metric of choice.

Typically, experiments lie somewhere in between wanting completely pure or completely efficient data sets, and so we can optimize a combination of these two metrics to evaluate challenge submissions. For example, the SNPhotCC figure of metric for deciding on classifier performance computed the product of the efficiency and what is called the `pseudo-purity' which included both classification purity and also selection effects based on the telescope performance. 

\subsubsection{Probabilistic and deterministic classifiers}
A great discriminant between different types of classifiers and classification challenges themselves, is whether deterministic or probabilistic classifications are computed. A deterministic classification produces a binary assignment of `True' or 'False' (or numerically a `1' or `0') for each class, based on some metric value, which can be computed at various threshold settings. Probabilistic classifications, however, give the expected probability $0\geq P_{ij}\leq 1$ that the object $i$ belongs to class $j$. If one considers an $i\times j$ classification matrix, sometimes referred to as a confusion matrix, then probabilistic classifiers will have all rows summing to unity such that the sum of probabilities over all classes is unity.

The probabilistic confusion matrix (this term is technically a misnomer in the probabilistic case, conditional probability matrix is more suited) can be converted into the standard deterministic `1's and `0's by assigning the deterministic type to the class with the highest probabilistic classification. The reverse process, where probabilities are computed from deterministic classifications, is not always as simple a procedure. 
\begin{figure}[htbp!]
    \centering
    \includegraphics[width=0.5\textwidth]{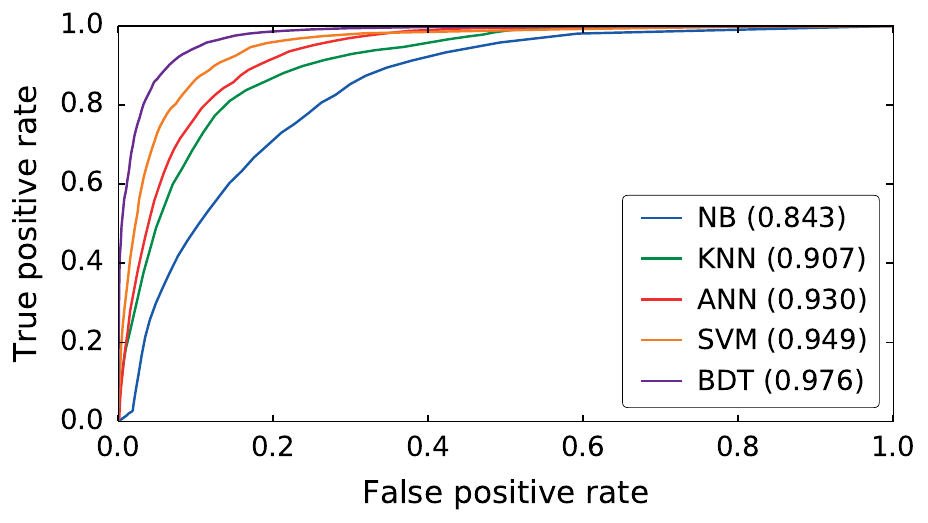}
    \caption{The Receiver Operating Characteristic (ROC) curve for SNPhotCC, which focused on classifying Type Ia supernovae problem according to various procedures, namely Naive Bayes classifiers (NB), K-nearest neighbors (KNN), artificial neural network (ANN), support vector machine (SVN) and boosted decision trees (BDT). `Good' classification approaches are close to right-angled in TP-rate--FP-rate space (having a high TP rate for all values of the FP rate). These ROC curves are computed for different classification thresholds. The area under these ROC curves, known as the AUC is interpreted as the probability that the classifier provides a higher score to a positive object than a negative object. Figure reproduced with permission from \cite{2016ApJS..225...31L}. 
    \label{fig:lochner}}
\end{figure}
In these cases, a few procedures can be followed. The first is to compute the receiver operating characteristic (ROC) curve, which shows the classifier performance by plotting the TP rate against the False Positive (FP) rate. The TP rate is the ratio of TP classifications to the sum of both the TP and FN classifications, or $ N_\mathrm{TP}/(N_\mathrm{TP}+N_\mathrm{FN}$). This is equivalent to the efficiency term defined previously. The FP rate is the ratio of FPs to the sum of FPs and TNss, or $N_\mathrm{FP}/(N_\mathrm{FP}+N_\mathrm{TN})$. ROC curves should be computed at a range of different classification thresholds, to accurately compute a classification probability. However, instead the area under that ROC curve, named the AUC for Area Under Curve, is then interpreted as the probability that the classifier scores a (random) positive object higher than a random negative object over all ``thresholds''. The AUC is normalized to lie between zero and unity. An example ROC curve is shown in Figure~\ref{fig:lochner}, from \cite{2016ApJS..225...31L}.

\subsubsection{Examples of classifiers}

The recent growth in classifier methodologies and types is partly due to the existence of new classification challenges and the data sets they are based on. The data sets themselves usually consist of light-curves: flux as a function of time. These light-curves are assumed to be derived based on the processing of images. Typically, difference images are created by subtracting two images taken on different nights and looking for bright `spots' of emission that has changed between the two images. The original images are then processed to determine the flux as a function of time for the object that was detected as varying in flux from the difference image.

Classification algorithms can broadly be classified into those performing supervised (or trained) learning about the object classes, or those employing unsupervised learning. 
In \textbf{supervised learning}, classification algorithms are trained by associating a set of characteristics, such as object properties, or the time-series variables themselves such as flux, with a set of truth labels for each object. The algorithm is then tested on a new or test data set to assess the quality of the classification algorithm, or how good the code was at learning which properties tracked the types well.
Supervised learning methods often include template-based approaches \citep[see e.g.][]{2018MNRAS.477.4142D, 2018MNRAS.473.3969R, 2016JCAP...12..008M, 2016ApJS..225...31L, 2016ApJ...817...73H, sesar, 2016MNRAS.455..626M, 2015MNRAS.453.2848V,  2013MNRAS.430..509I, 2007AJ....134.1285P}. 
Template-based methods allow for the inclusion of physical information priors on the types of objects, and allow potentially one to use the classification output to learn about which physical parameters correlate with classification.

Supervised learning is however very dependent on the quality of the training data set, and how representative it is of the test data. Non-representativity (the inability of the training data to reflect the properties of the future test data) if unaccounted for, can lead to extremely poor performance of supervised learning algorithms. Unfortunately, given the drive of the astronomical community to push to greater depths, redshifts and areas on the sky, the (future/test) data are almost guaranteed to be different from the data in hand that are typically used to train algorithms.
\begin{figure*}[htbp!]
    \centering
    \includegraphics[width=0.95\textwidth]{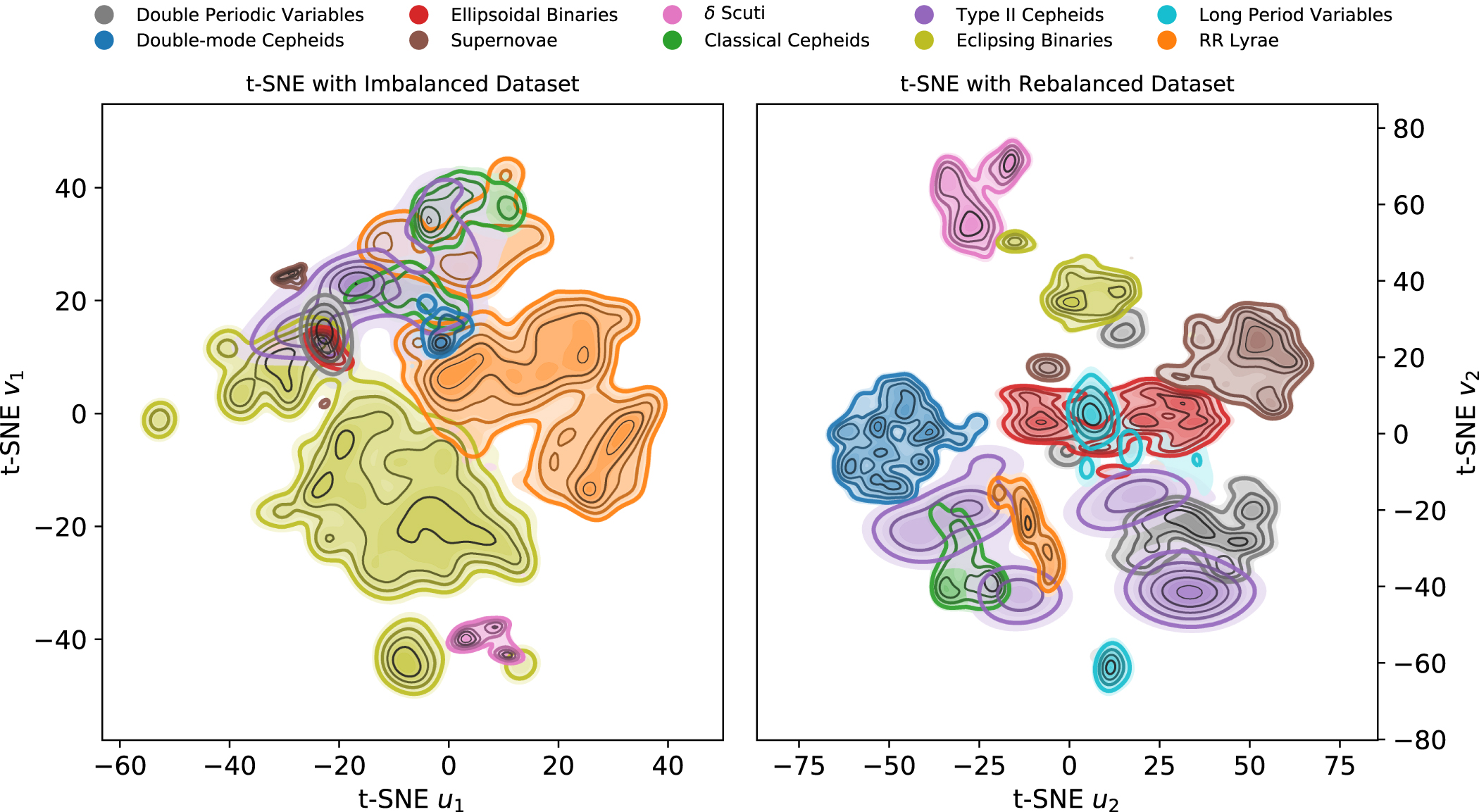}
    \caption{Clustering of principal components in a hybrid classification scheme: the ANTARES transient `broker' or classifier, performs a principal component analysis on the light-curve features, uses the t-Distributed Stochastic Neighbor Embedding (t-SNE) to reduce the dimensionality of the space and addresses class imbalance that results from the non-representativity of the training data. Figure reproduced with permission from \cite{2018ApJS..236....9N}. 
    \label{fig:antares}}
\end{figure*}
A more agnostic approach is to perform \textbf{unsupervised learning}. In unsupervised learning, one does not make use of any labels, and instead infers directly the structure of properties in the data set. Examples of unsupervised learning include clustering, representation learning and density estimation. This is useful to identify classes themselves, and are most useful to determining how many classes there are in the data set, with the aim to reducing the dimensionality of a data set as it becomes clear which properties are, or are not, informative of the different classes in the data.

Often, hybrid semi-supervised learning methods are used to improve performance over either individual method, by computing general features or performing a decomposition over feature space (see e.g. \cite{2018ApJS..236....9N}, and the figure reproduced as Figure~\ref{fig:antares}) but makes use of labels if they are available. Similarly, classifiers typically operate either on the light-curve (time series) data that originate from processing images \citep[see e.g.][]{2012MNRAS.419.1121R, 2011MNRAS.414.1987N} or work directly on images. Some classification approaches use image-based techniques \citep[e.g. neural networks on transformed images from time-series data][]{2017arXiv170906257M, 2017ApJ...837L..28C}.



\subsection{Anomaly detection}
In contrast to the problem of classification of known types of objects in new regimes and with new facilities, some of the greatest scientific challenges and areas of interest are to detect and understand anomalies in astronomical data. New telescopes with fast survey speed will generate large volumes of data. In addition to wanting a pure sample of any particular type of object, one also does not want to waste spectroscopic follow-up resources on objects that were incorrectly classified. Hence, studies of simulated data where one knows the true object classes are key to asses both the purity of a sample and efficiency of classification. This is especially true for rare transients where follow up resources need to be triggered as soon as possible. 

This active area of research \citep[see e.g.][]{Xiong2010AnomalyDF, Henrion2013, 2014ApJ...793...23N, doi_10.1117/12.2231491, 2018arXiv181108055Z} typically identifies new celestial objects of interest. Some groups are interested in finding new objects of interest in a set, however \cite{Xiong2010AnomalyDF} discuss the separation of anomalies into point-like anomalies or objects and group-like anomalies, and discuss methods for detecting those types separately.

Point-like anomaly detection occurs when a large and often high-dimensional data set must be divided into sub-spaces, and the features of any object in the space can be reconstructed from a linear basis of features, through methods like Mixed-Error Matrix Factorization, an approach advocated in \cite{Xiong2010AnomalyDF}). Anomalies are therefore detected as those that are not well reconstructed from these basis functions. In the case of group anomaly detection, the group is treated as made up of exchangeable members, so each member is expressed as a mixture of all possible classes. In this case hierarchical modeling can be used to characterize the group as a whole and a score of ``anomalousness'' can be derived for each member \cite{Xiong2010AnomalyDF}. This has the added benefit that while searching for anomalies, one can also learn about the group characteristics themselves.  Bayesian approaches are well suited here. One example is Bayesian Anomaly Detection And Classification \citep[BADAC,][]{2019arXiv190208627R}, that computes the probabilities that a new measurement belongs to each of a set of known classes while simultaneously ranking the objects by this anomalousness.

\begin{figure*}[htbp!]
    \begin{center}
    \includegraphics[ width=0.85\textwidth]{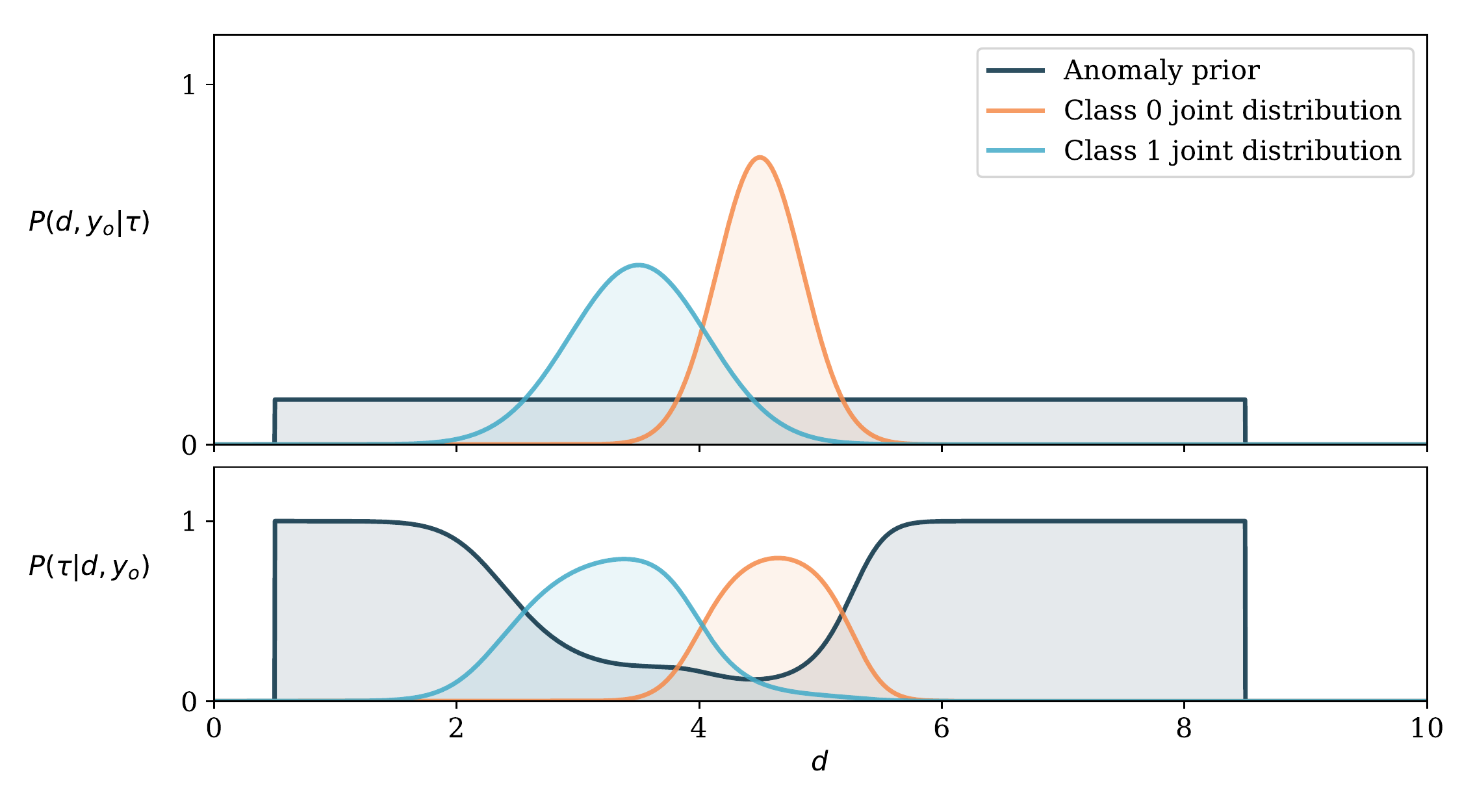}
    \caption{The schematic summary of Bayesian Anomaly Detection And Classification (BADAC). The top panel shows the two joint distributions for the test value $d$, with a flat prior for `unknown' anomalies. The bottom panel shows the posterior class values for changing $d$. The posterior probability for being an anomaly is the lowest where there is support from the known classes, but increases over the range of $d$ where there is no training data. Figure reproduced with permission from \cite{2019arXiv190208627R}.}
    \end{center}
\end{figure*}

While time-series data are perhaps most suited to the detection of anomalies, the search for anomalies is applied to broader problems in astronomy like photometric redshift anomalies \citep[see][as one example.]{2015MNRAS.452.4183H}

\section{Logistics of Time-series challenges}
In order to bring a time-series data challenge to fruition requires a significant amount of effort, in the preparation and validation of the data, and in determining the metrics to use in order to evaluate entries to the challenge, once it is underway. 

\subsection{Data preparation}
Time-domain challenges have become avenues to engage communities that contain those who are experts and non-experts in astronomy. This broad appeal creates additional tasks for those preparing a data challenge. Once models (or existing data) are collated to form part of the data challenge, these data need to be calibrated and validated, in order to ensure that the data are robust to `flags' inserted (e.g. making sure a simulation is randomized, or requiring that the telescope flag for a given observation does not act as an addition label for the type of object).

As a first step, one validates that the simulations are a faithful representation of the input models, or historical data. For a time-series challenge like PLAsTiCC, this involved plotting the theoretical model without noise over any real-world examples of the class, and comparing the model with the noise simulation of the same model. This provided a useful cross-check of any spurious artifacts in the simulation data. For the PLAsTiCC challenge, this validation effort is described in \cite{validation}, while the models will be described in \cite{modelling}.

For a simulation coming from one underlying model rather than an assortment, one must show that the model considered can reproduce the variety of simulated data that match observations. In the GREAT3 challenge, this amounted to confirming that GalSim\footnote{\url{https://github.com/GalSim-developers/GalSim}} \citep{galsim} was indeed able to produce both realistic sheared images galaxies and known parametric (e.g. S\'ersic) models, and that the simulations could be used to test different weak lensing algorithms. In this case \cite{great3} compared the estimated shear from the simulations as measured by traditional `photon shooting' methods and compared them to the discrete Fourier transform approach used in Galsim to show consistency in the simulations. After the consistency checks, the Galsim suite was used to simulate a large range of galaxies that was trained on Hubble Space Telescope (HST) data, but that was extended to fainter samples for the challenge.
For the Strong Gravitational Lens Finding Challenge the simulations were based on large scale structure simulations in which dark matter halos were identified, and ray tracing was used to lens foreground galaxy images which were generated to mimic both ground-based multi-band surveys, like the Kilo-Degree Survey \citep[KIDS\footnote{\url{http://kids.strw.leidenuniv.nl/}},][]{2013ExA....35...25D} and future space-based surveys like Euclid\footnote{\url{http://sci.esa.int/euclid/}}. Realistic point spread functions (PSF) were used to match current surveys, hence overall image properties can be validated off existing strong lens data.

\subsection{Evaluation Metrics}
The criteria for evaluating different classification schemes depend on the goals of the particular time-series challenge. Some challenges and collaborations have favored some purity of classification of one type of object over efficiency and classification challenges grow in complexity to match the expected complexity of the data set. For upcoming heterogeneous surveys like LSST, however, that approach would result in a splinter of metrics for each individual science case/object of interest.
The approach of the PLAsTiCC team was to employ a metric that explicitly favored classification over the entire set of classes \cite{2018arXiv180911145M}. This led to the choice of log-loss as the metric:

\begin{eqnarray}
  \label{eq:logloss}
  L_{n} &\equiv & -\sum_{m=1}^{M} \tau_{n, m} \ln[p(m \mid d_{n})],
\end{eqnarray}

where \begin{eqnarray}
  \label{eq:indicator}
  \tau_{n, m} &\equiv& 
  \begin{cases}
  0 & n \neq m \\
  1 & n = m
  \end{cases}
\end{eqnarray}
is the matrix that evaluates the metric over the true classes, $p(m \mid d_n)$ is the posterior probability of the $n$-th object is of class $m$. This metric is additionally weighted over classes, so that one is not rewarded for correctly classifying the most populous classes correctly, but that the classification algorithm must classify all objects evenly within a specified level of accuracy.

The Strong Gravitational Lens Finding Challenge was also focused on classifying identifying strong lens systems in simulated images. In this case metric used to evaluate classification submissions to the challenge was the area under the receiver operating curve (AUROC). Given how rare strong lens systems are, however, the rate of lenses in the training set was boosted in order to facilitate accurate  classification, and hence the absolute number of FPs (contamination) will be enhanced relative to the real data. To account for this, those running the challenge included the highest True Positive Rate (TPR$_0$) as a label in the classification scheme. The TPR$_0$ is the highest probability $p$ before a single false positive occurs in the test set of 100,000 systems. Similarly, TPR$_{10}$ is defined as the highest TPR before ten FPs are recorded. These metrics are defined to make contact with future observational requirements for purity, given the cost of following up strong lens systems with additional facilities.



Challenges based on computation of a result rather than classification (e.g. the GREAT3 challenge), the metrics defined depended on specific conditions of the data. For example, one metric was used applied to galaxies that had been simulated with constant shear branches:

\begin{equation}
Q_c = \frac{2000\times \eta_c}{\sqrt{\sigma^2_\mathrm{min,c}}+\sum_{i=+, \times} \left(\frac{m_i}{m_\mathrm{target}}\right)^2+\left( \frac{c_i}{c_\mathrm{target}}  \right)^2} 
\end{equation}
where the $m$ and $c$ are the multiplicative and additive bias respectively. For GREAT3 the target values were $m_\mathrm{target} = 2\times10^{-3}, c_\mathrm{target} = 2\times10^{-4}.$ The normalization factor $\eta_c=1.232$ was used in the challenge, and the minimum variance assumed for ground (space) challenges was $\sigma_\mathrm{min,c}^2 = 4 (1).$

\begin{figure*}[htbp!]
    \begin{center}
    \includegraphics[, angle=270, width=0.85\textwidth]{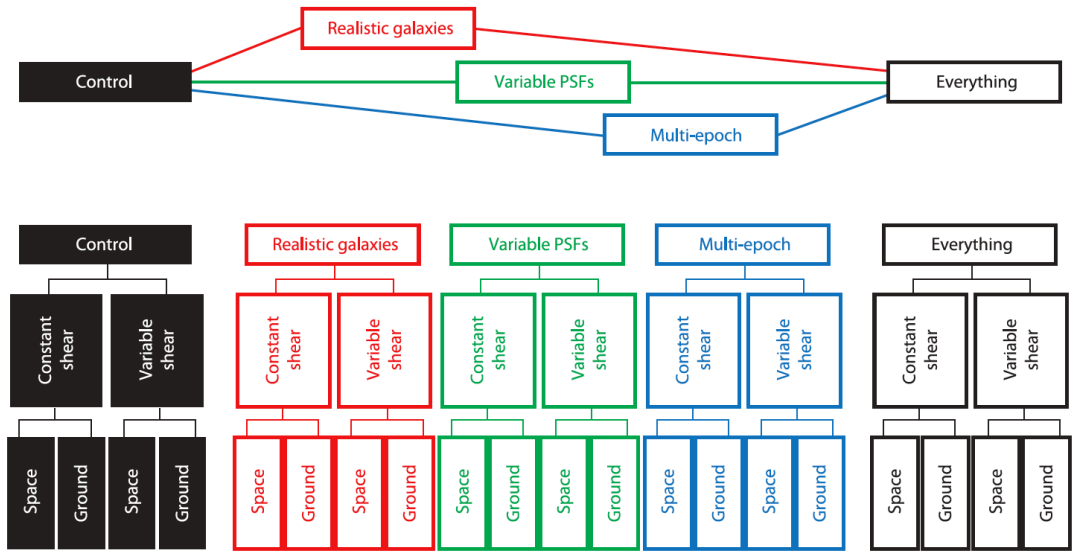}
    \caption{The GREAT3 data and `branch' structure. Participants were asked to compute the shear signal on simulated galaxies, and their performance was evaluated on different branches depending on whether a constant or a variable shear had been applied to the simulated data. Figure reproduced with permission from \cite{2015A&C....10..121R}}.
    \end{center}
\end{figure*}

 \section{Future data challenges as tools for new methods}
  Data challenges for time-series data will continue to provide one of the best avenues for generating new methods of classification and anomaly detection. Several avenues exist for extension and broadening of the science goals and aims of such challenges. Some challenges exist in the catalog domain, by assuming difference images have been analyzed to obtain fluxes as a function of time for the objects of interest.
  While this has long been the case for studies involving gravitational lenses \cite{2018arXiv180203609M}, future iterations of e.g. the PLAsTiCC time-series challenge will be focused on image data directly, with the inclusion of more information on the potential galaxy hosting the variable or transient object.
 
Some of the issues faced by those preparing astronomy challenges to the community include mapping a given problem in astronomy to one metric, as seen in the strong lensing challenge and the weighted metric of PLAsTiCC. While some challenges mitigate this with multiple metrics, others include various stages of classification (e.g. GalaxyZoo) to ensure information is gathered to enable studies of systematics and degeneracies along the classification process. 

Large, public challenges provide a framework for testing a host of approaches, however the heterogeneity of classification approaches needs to be considered to ensure that a given participation metric doesn't penalize a given classification approach. 

Challenges that live in the `live streaming' regime are essential to test the fidelity of classification for fast, wide surveys that will deliver a great deal of real-time data. This is something of great relevance to current experiments in the optical like ZTF and also in the radio like CHIME, who are already performing classification and anomaly detection on data streams \textit{in situ}.
 
Common problems exist for the processing of astronomical data from completely different wavelength regimes, however there are often barriers to entry to collaboration across different fields and science interests. Preparing tools and software that are able to absorb different types of data remains a challenge to the astronomy community if we are to learn the best practice for future challenges.

\section{Summary}
The large, complex and multidimensional data sets in astronomy provide an excellent test bed for new statistical tools and methodologies. New discoveries will be possible with these new data, and the ability to learn about objects using observations across a wide range or wavelengths has the potential to uncover the underlying physical process involved in the transient objects that are temporary guests in the sky. Data challenges from either real or simulated time-domain data are powerful tools to not only learn about the data sets expected from ambitious new projects like LSST and the SKA, but also act as beacons to methodology development by challenging the community to address problems of sparse, imbalanced and heterogeneous data.  

\section{acknowledgements}
I wish to thank Ashish Mahabal for useful discussions about this review, and the PLAsTiCC team for their astronomical work on our recent data challenge. PLAsTiCC was supported by the LSST Corporation. This work is supported by Natural Sciences and Engineering Research Council of Canada. The Dunlap Institute is funded through an endowment established by the David Dunlap family and the University of Toronto.
I acknowledge that the land on which the University of Toronto is built is the traditional territory of the Haudenosaunee, and most recently, the territory of the Mississaugas of the New Credit First Nation. I am grateful to have the opportunity to work in the community, on this territory.




    

\bibliographystyle{aasjournal}
\bibliography{challenge}




\end{document}